\documentclass[11pt]{JHEP3} 

\usepackage{epsfig,axodraw,here}

\newcommand{\be}[1]{\begin{equation} \label{(#1)}}
\newcommand{\ee}{\end{equation}}
\newcommand{\baq}[1]{\begin{eqnarray} \label{(#1)}}
\newcommand{\eaq}{\end{eqnarray}}
\newcommand{\nn}{\nonumber}
\newcommand{\rf}[1]{(\ref{(#1)})}

\def\rp{$R_p \hspace{-1em}/\;\:$ }

\newcommand {\ignore}[1]{}

\def\nn{\nonumber}
\def\rp{$R_p \hspace{-1em}/\;\:$ }

\def\bear{\be\begin{array}}
\def\eear{\end{array}\ee} 
\def\bea{\begin{eqnarray}} 
\def\eea{\end{eqnarray}}

\def\vb#1{\vbox to #1 pt{}}
\def\beqa{\begin{eqnarray}}
\def\eeqa{\end{eqnarray}}

\def\beq{\begin{equation}}
\def\eeq{\end{equation}}
\def\ba{\begin{array}}
\def\ea{\end{array}}

\def\ds{\displaystyle}

\def\21{$SU(2) \otimes U(1)$}

\def\half{{\textstyle{1 \over 2}}}

\def\eighth{{\textstyle{1 \over 8}}} 
 
\def\bold#1{\setbox0=\hbox{$#1$} 
     \kern-.025em\copy0\kern-\wd0 
     \kern.05em\copy0\kern-\wd0 
     \kern-.025em\raise.0433em\box0 } 
\newcommand{\phim}{\varphi_{\mu}}
\newcommand{\phiee}{\varphi_{\epsilon_1}}
\newcommand{\phiem}{\varphi_{\epsilon_2}}
\newcommand{\phiet}{\varphi_{\epsilon_3}}
\newcommand{\phiei}{\varphi_{\epsilon_i}}
\newcommand{\phiBe}{\varphi_{B_1}}
\newcommand{\phiBm}{\varphi_{B_2}}
\newcommand{\phiBt}{\varphi_{B_3}}
\newcommand{\phiBi}{\varphi_{B_i}}

\newcommand{\phiB}{\varphi_{B}}

\title{\large CP violation in decays of the lightest
 supersymmetric particle with bilinearly broken R parity}

\author{M. Hirsch ${}^{a}$, T. Kernreiter ${}^{a,b}$
\\ 
${}^{a}$ Astroparticle and High Energy Physics Group, 
IFIC - Instituto de F\'\i sica Corpuscular,
Edificio Institutos de Investigaci\'on,
Apartado de Correos 22085,
E-46071 Valencia - Spain\\
${}^{b}$ Institut f\"ur Theoretische Physik, Universit\"at Wien, \\ 
A-1090 Vienna, Austria}

\author{W. Porod
\\ Institut f\"ur Theoretische Physik, Universit\"at Z\"urich, \\ 
CH-8057 Z\"urich, Switzerland}

\abstract{
Supersymmetric models with broken R-parity induced by lepton number
violating terms provide a calculable framework for neutrino masses and
mixings.  Within models with bilinear R-parity breaking six new
physical phases appear which are potential sources of novel
CP-violating phenomena compared to the minimal supersymmetric
extension of the standard model. 
We consider CP-violating observables in the decays
of the lightest supersymmetric
particle in this class of models. We show that: (i) Neutrino physics
requires a strong correlation between three different pairs of phases,
thus reducing the effective number of new phases to three. (ii)
CP-violating phenomena in decays of the lightest supersymmetric
particle due to new R-parity breaking phases turn
out to be small, once constraints from neutrino physics are taken
into account. We demonstrate that this feature does not depend on the
nature of the lightest supersymmetric particle.}

\keywords{supersymmetry, R parity, violation, violation, CP}

\preprint{hep-ph/0211446\\  IFIC/02-57\\  ZU-TH 22/02 \\
   UWThPh-2002-35}

\begin{document}

\section{Introduction}

The standard model (SM) contains a source for CP violation, the 
complex phase of the Kobayashi-Maskawa matrix \cite{Kobayashi:fv}. 
However, it has often been argued that the SM is not able to explain the 
observed baryon/antibaryon asymmetry of the universe (see for example 
\cite{BGSM}) and thus new physics might be expected to show up in CP 
violating phenomena. 

In supersymmetric (SUSY) extensions of the SM, with their much larger particle 
content, a considerable number of parameters could in principle
be complex and consequently CP violation phenomena differ 
considerably from SM expectations \cite{Masiero:xj}.
It is therefore not surprising that the study of CP violation in the 
minimal supersymmetric extension of the standard model (MSSM) 
\cite{Haber:1984rc} or its constrained version (sometimes called 
CMSSM or mSUGRA in the literature) has received quite some attention 
recently \cite{CPVMSSM,CPVMSSM1}.

The new phases present in SUSY models, however, can be 
restricted by existing upper limits on electric dipole moments 
\cite{Masiero:xj}. The general consensus is that (at least) one of 
the following three conditions has to be realized: (i) The phases 
are severely suppressed \cite{Falk:1996ni}. 
(ii) Supersymmetric particles of the first two generations are 
rather heavy, with masses of the order of a few TeV at least 
\cite{Dimopoulos:1995mi}. (iii) There 
is a rather strong correlation between phases \cite{cancel}, 
leading to a cancellation of the different SUSY contribution 
to EDMs \footnote{The correlation ``solution'' is somewhat debated, 
see for example  \cite{Falk:1996ni}.}.

Far less work, however, has been devoted up to now to CP violation 
in supersymmetry with broken R-parity (\rp)~\cite{cpvrpv}. R-parity 
breaking implies a violation of either lepton or baryon number. 
It is phenomenologically unacceptable that both types of 
terms are present \cite{Dreiner:1997uz}. We will discuss only the 
lepton number violating terms in the following, because
we focus on connections to neutrino physics.

R-parity can be broken by bilinear and by trilinear terms. If both 
types were present up to $(36+3)\times 2$ new parameters appear 
in the theory, all of which could be complex. {\em A priori} one could 
therefore expect that CP violating phenomena differ considerably 
from both the SM as well as the MSSM. 

Electric dipole moments do not constrain in a significant way 
the phases of either trilinear or bilinear terms individually, 
essentially since the leading contributions to the EDMs occur only at 
2-loop level, if either only trilinear or only bilinear terms are present 
\cite{2loopEDM}.~\footnote{1-loop contributions are proportional 
to the neutrino mass, see the first paper in \cite{2loopEDM}.} 
If both types of R-parity breaking terms were 
present contributions to EDMs appear at 1-loop level \cite{Keum:2000ea} 
and thus the imaginary part of a certain product of bilinear and 
trilinear terms is more tightly constrained. However, even in the 
latter case the limit is not especially strong, considering the 
typical size of R-parity violating parameters expected from neutrino 
physics.
It is thus fair to say that very little is known currently about 
the phases of R-parity breaking parameters.

In this paper we study CP violation in SUSY with bilinear 
R-parity breaking, focusing mainly on aspects of those phases which 
are not present in the MSSM. The assumption of having only bilinear 
R-parity violation reduces the number of new phases to only six, as 
will be discussed in more detail later in the paper. 

Studies of bilinear R-parity breaking SUSY at this moment are mainly 
motivated by the recent discoveries in neutrino physics. Observations 
of atmospheric neutrinos by the Super-K collaboration 
\cite{Fukuda:1998mi} have confirmed the deficit of atmospheric muon
neutrinos, especially at small zenith angles, and thus strongly 
point to non-zero neutrino masses and mixings. The preferred range 
of oscillation parameters from atmospheric neutrino data is currently 
(at 3$\sigma$ and 1 d.o.f.) \cite{Fukuda:1998mi,nu2002} 
\begin{equation}
\label{eq:atm}
    0.3 \le \sin^2\theta_{Atm} \le 0.7 \,,\quad
    1.2 \times 10^{-3}~{\rm eV^2} \le   \Delta m^2_{Atm}\le 4.8 \times
    10^{-3}~{\rm eV^2} \:.
\end{equation}
Also the long-standing solar neutrino problem now provides strong 
evidence for neutrino flavour conversion, especially considering 
the recent measurement of the neutral current rate for solar neutrinos 
by the SNO collaboration \cite{Ahmad:2002jz}. If interpreted in terms 
of neutrino oscillations, the data indicate a large mixing angle 
between $\nu_e$ and $\nu_{\mu}-\nu_{\tau}$, with a strong preference 
towards the large mixing angle MSW solution (LMA).  At $3\sigma$ one
has~\cite{nu2002,Maltoni:2002ni}
\begin{equation}
\label{eq:solbound}
0.25 \le \tan^2\theta_{\odot} \le 0.83  
\end{equation}
for 1 d.o.f., the best-fit-parameters being 
\begin{equation}
\label{eq:solbfp}
  \tan^2\theta_{\odot} = 0.44 \:\:\:\: 
\Delta m^2_{\odot} = 6.6\times 10^{-5}~{\rm eV^2}
\end{equation}
This nicely confirms earlier hints found in Ref.~\cite{Solar}.

Calculated at the 1-loop level \cite{Hempfling,NuMass},
SUSY with bilinear R-parity breaking 
can explain the solar and atmospheric 
neutrino data \cite{NuMass}. However, \cite{Hempfling,NuMass} 
considered only real parameters and thus were not concerned about 
possible CP-violating effects. 

We will essentially follow \cite{NuMass}, extending the calculation 
to the complex case. In the next section we will give mass matrices 
and the Higgs potential of the model. In Section 3 we will discuss 
the constraints on the various phases of the model implied by 
current neutrino data. Section 4 will discuss possible CP-violating 
observables, before we close with a short conclusion. 

\section{The Model}

\subsection{Superpotential and soft SUSY breaking}

The supersymmetric Lagrangian is specified by the 
superpotential $W$ 
\be{SuperPot}
W=\varepsilon_{ab}\left[ 
 h_U^{ij}\widehat Q_i^a\widehat U_j\widehat H_u^b 
+h_D^{ij}\widehat Q_i^b\widehat D_j\widehat H_d^a 
+h_E^{ij}\widehat L_i^b\widehat R_j\widehat H_d^a 
-\mu\widehat H_d^a\widehat H_u^b 
+\epsilon_i\widehat L_i^a\widehat H_u^b\right]~, 
\ee 
where $i,j=1,2,3$ are generation indices, $a,b=1,2$ are $SU(2)$
indices, and $\varepsilon$ is the completely antisymmetric $2\times2$
matrix, with $\varepsilon_{12}=1$. The symbol ``hat'' over each letter
indicates a superfield, with $\widehat Q_i$, $\widehat L_i$, $\widehat
H_d$, and $\widehat H_u$ being $SU(2)$ doublets with hypercharges
$\frac{1}{3}$, $-1$, $-1$, and $1$, respectively, and $\widehat U$,
$\widehat D$, and $\widehat R$ being $SU(2)$ singlets with
hypercharges $-{\textstyle{4\over 3}}$, ${\textstyle{2\over 3}}$, and
$2$, respectively. The couplings $h_U$, $h_D$ and $h_E$ are $3\times 3$
Yukawa matrices, and $\mu$ and $\epsilon_i$ are parameters with units
of mass. 

Supersymmetry breaking is parameterized with a set of soft 
supersymmetry breaking terms,
\baq{Vsoft}
V_{soft}&=& 
M_Q^{ij2}\widetilde Q^{a*}_i\widetilde Q^a_j+M_U^{ij2} 
\widetilde U_i\widetilde U^*_j+M_D^{ij2}\widetilde D_i 
\widetilde D^*_j+M_L^{ij2}\widetilde L^{a*}_i\widetilde L^a_j+ 
M_R^{ij2}\widetilde R_i\widetilde R^*_j \nonumber\\ 
&&\!\!\!\!+m_{H_d}^2 H^{a*}_d H^a_d+m_{H_u}^2 H^{a*}_u H^a_u- 
\left[\half M_3\lambda_3\lambda_3+\half M_2\lambda_2\lambda_2 
+\half M_1\lambda_1\lambda_1+{\rm h.c.}\right] \nn \\ 
&&\!\!\!\!\!\!\!\!\!\!\!\!\!\!\!\!\!\!\!\!+\varepsilon_{ab}\left[ 
A_U^{ij}\widetilde Q_i^a\widetilde U_j H_u^b 
+A_D^{ij}\widetilde Q_i^b\widetilde D_j H_d^a 
+A_E^{ij}\widetilde L_i^b\widetilde R_j H_d^a 
-B\mu H_d^a H_u^b+B_i\epsilon_i\widetilde L_i^a H_u^b+{\rm h.c.}\right] 
\eaq
In the following we assume that there is no intergenerational mixing in 
the soft terms. Let us first list the parameters which may be 
complex in the model defined by Eq.~\rf{SuperPot} and 
Eq.~\rf{Vsoft}. 
Decomposed in modulus and phase these are given by
$\epsilon_i=|\epsilon_i|e^{i\phiei}$ and 
$\mu=|\mu|e^{i\phim}$ in 
Eq.~\rf{SuperPot}, and 
$A_U={\rm diag}\{|A_u|e^{i\varphi_{A_u}}$, 
$|A_c|e^{i\varphi_{A_c}}$, 
$|A_t|e^{i\varphi_{A_t}}\}$, 
$A_D={\rm diag}\{|A_d|e^{i\varphi_{A_d}}$, 
$|A_s|e^{i\varphi_{A_s}}$, 
$|A_b|e^{i\varphi_{A_b}}\}$, 
$A_E= {\rm diag}\{|A_e|e^{i\varphi_{A_e}}$,
$|A_{\mu}|e^{i\varphi_{A_{\mu}}}$, 
$|A_{\tau}|e^{i\varphi_{A_{\tau}}}\}$, 
$M_1=|M_1|e^{i\phi_1}$, 
$M_2=|M_2|e^{i\phi_2}$, 
$M_3=|M_3|e^{i\phi_3}$, 
$B\mu=|B\mu|e^{i\phiB}$ and
$B_i\epsilon_i=|B_i\epsilon_i|e^{i\phiBi}$
in Eq.~\rf{Vsoft}.
This means, in addition to the MSSM parameters
the following six parameters are in general complex.
\be{Rphases}
\epsilon_i=|\epsilon_i| e^{i\phiei}\, ,\qquad 
B_i\epsilon_i=|B_i\epsilon_i|e^{i\phiBi}\, .
\ee
As mentioned no restrictions on the size of
the phases in Eq.~\rf{Rphases} exist;
they can be $O(1)$.

At this point it is appropriate to note that not all of the
phases quoted above have a physical meaning.
Any two of these CP--odd phases may be eliminated by
employing the Peccei--Quinn- and the $R$ symmetry
$U(1)_{PQ}$ and $U(1)_R$, respectively. We will use this 
phase freedom later on to remove two unphysical phases.

\subsection{Scalar potential}

Next we consider the scalar potential and derive
the tadpole equations.
The scalar potential at tree level is
\begin{equation}
\label{Vtotal}
V = \sum_i \left| { \partial W \over \partial z_i} \right|^2 
        + V_D + V_{soft}.
\end{equation} 
where $z_i$ is any one of the scalar fields in the superpotential in
Eq.~\rf{SuperPot}, $V_D$ are the $D$-terms, and $V_{soft}$ is
given in Eq.~\rf{Vsoft}. 
The vacuum expectation values
(vevs) of the Higgs ($H_u, H_d$) and lepton ($\tilde{L}_i$)
fields are complex in general.
In order to determine the ground state of the potential
Eq.~(\ref{Vtotal}), we use the linear parametrization
\begin{eqnarray} 
H_d &=& e^{i\theta}{{\frac{1}{\sqrt{2}}[\sigma^0_d+v_d+i\varphi^0_d]}
\choose{H^-_d}}\,,\qquad
H_u={{H^+_u}\choose{\frac{1}{\sqrt{2}}
[\sigma^0_u+v_u+i\varphi^0_u]}},\nonumber \\[2mm]
&&
\widetilde L_i=e^{i\eta_i}{{\frac{1}{\sqrt{2}}[\tilde\nu^R_i+
v_i+i\tilde\nu^I_i]}\choose{\tilde\ell^-_i}}\,,
\label{eq:shiftdoub} 
\end{eqnarray} 
where $v_d, v_u$ and $v_i$ are positive,
and we have set the phase of $H_u$ to zero since only
relative phases are meaningful.
The stationary conditions then read
\baq{Tadeven}
\partial\langle V\rangle\over{\partial v_d}
&=&\Big(m_{H_d}^2+|\mu|^2\Big)v_d+v_dD-
|B\mu| v_u \cos(\phiB+\theta)-v_i|\mu||\epsilon_i|
\cos(\phim+\theta-\phiei -\eta_i)=0
\nonumber\\
\partial\langle V\rangle\over{\partial v_u}
&=&\Big(m_{H_u}^2+|\mu|^2+|\epsilon_i|^2\Big)v_u-
v_uD+v_i|B_i\epsilon_i|\cos(\phiBi+\eta_i)-
|B\mu| v_d\cos(\phiB+\theta)=0
\nonumber\\
\partial\langle V\rangle\over{\partial v_1}
&=&v_1D+|\mu||\epsilon_1| v_d\cos(\phim+\theta-\phiee-\eta_1)+
v_1\Big(|\epsilon_1|^2+M^2_{L_1}\Big)+
|B_1\epsilon_1|v_u\cos(\phiBe+\eta_1)\nonumber \\
&&+v_2\epsilon_2\epsilon_1
\cos(\phiee+\eta_1-\phiem-\eta_2)
+v_3\epsilon_3\epsilon_1
\cos(\phiee+\eta_1-\phiet-\eta_3)=0
\nonumber\\
\partial\langle V\rangle\over{\partial v_2}
&=&v_2D+|\mu||\epsilon_2| v_d\cos(\phim+\theta-\phiem-\eta_2)+
v_2\Big(|\epsilon_2|^2+M^2_{L_2}\Big)+
|B_2\epsilon_2|v_u\cos(\phiBm+\eta_2)\nonumber \\
&&+v_1\epsilon_1\epsilon_2
\cos(\phiem+\eta_2-\phiee-\eta_1)
+v_3\epsilon_3\epsilon_2
\cos(\phiem+\eta_2-\phiet-\eta_3)=0
\nonumber\\
\partial\langle V\rangle\over{\partial v_3}
&=&v_3D+|\mu||\epsilon_3| v_d\cos(\phim+\theta-\phiet-\eta_3)+
v_3\Big(|\epsilon_3|^2+M^2_{L_3}\Big)+
|B_3\epsilon_3|v_u\cos(\phiBt+\eta_3)\nonumber \\
&&+v_1\epsilon_1\epsilon_3
\cos(\phiet+\eta_3-\phiee-\eta_1)
+v_2\epsilon_2\epsilon_3
\cos(\phiet+\eta_3-\phiem-\eta_2)=0
\eaq
where $D=\eighth(g^2+g'^2)(v_1^2+v_2^2+v_3^2+v_d^2-v_u^2)$,
and
\baq{Tadodd}
\partial\langle V\rangle\over{\partial \theta}
&=&v_u v_d |B\mu|\sin(\phiB+\theta)+v_i v_d|\epsilon_i||\mu|
\sin(\phim+\theta-\phiei -\eta_i)=0
\nonumber\\
\partial\langle V\rangle\over{\partial \eta_1}
&=&v_1 v_d |\mu||\epsilon_1| \sin(\phim+\theta-\phiee-\eta_1)+
v_1 v_u |B_1\epsilon_1|\sin(\phiBe+\eta_1)\nonumber \\
&&+v_1 v_2\epsilon_1\epsilon_2
\sin(\phiee+\eta_1-\phiem-\eta_2)
+v_1 v_3\epsilon_1\epsilon_3
\sin(\phiee+\eta_1-\phiet-\eta_3)=0
\nonumber\\
\partial\langle V\rangle\over{\partial \eta_2}
&=&v_2 v_d |\mu||\epsilon_2| \sin(\phim+\theta-\phiem-\eta_2)+
v_2 v_u |B_2\epsilon_2|\sin(\phiBm+\eta_2)\nonumber \\
&&-v_2 v_1\epsilon_2\epsilon_1
\sin(\phiee+\eta_1-\phiem-\eta_2)
+v_2 v_3\epsilon_2\epsilon_3
\sin(\phiem+\eta_2-\phiet-\eta_3)=0
\nonumber\\
\partial\langle V\rangle\over{\partial \eta_3}
&=&v_3 v_d |\mu||\epsilon_3| \sin(\phim+\theta-\phiet-\eta_3)+
v_3 v_u |B_3\epsilon_3|\sin(\phiBt+\eta_3)\nonumber \\
&&-v_3 v_1\epsilon_3\epsilon_1
\sin(\phiee+\eta_1-\phiet-\eta_3)
-v_3 v_2\epsilon_2\epsilon_3
\sin(\phiem+\eta_2-\phiet-\eta_3)=0
\nonumber\\
\eaq
where the repeated index $i$ implies
a summation over $i=1,2,3$.

It was shown in Ref.~\cite{masip} that spontaneous CP 
violation does not occur in this model with $v_i\neq 0$, see also
\cite{joshipura}.
In the case of explicit CP violation, where 
$\{\phim,\phiB,\phiei,\phiBi\}\neq 0,\pi$,
the vevs are complex in general, 
as can be seen from Eq.~\rf{Tadodd}.

\subsection{Scalar mass matrices}

With the solutions to Eq.~\rf{Tadeven} and \rf{Tadodd}
we can give the mass matrices of the matter fields.
The mass matrix for the neutral boson fields
is a $10\times 10$ symmetric matrix.
The corresponding Lagrangian is  
\be{Neutmass} 
{\mathcal L}^{S}_{\rm mass}=
-\half S^T {\bold M^2_S} S, 
\ee 
where we have defined
$S^T\equiv[\sigma^0_d,\sigma^0_u,\tilde\nu_{i}^R,
\varphi^0_d,\varphi^0_u,\tilde\nu_{i}^I]$.
The mass matrix in Eq.~\rf{Neutmass} can be decomposed
into three $5\times 5$ blocks:
\be{MS}
{\bold M_{S}^2} =
\left[\matrix{
{\bold M_{SS}^2} &  {\bold M_{SP}^2} \cr
\vb{16}
{\bold M_{SP}^T}&
{\bold M_{PP}^2}  
}\right]
\ee
This mass matrix is diagonalized through
\be{Rotmat}
R {\bold M_{S}^2} R^T=
{\rm diag}(0,m^2_{S_1},{}_{\cdots},m^2_{S_9}),
\ee
where $R$ is a orthogonal matrix.
We give here only the scalar--pseudoscalar
mixing block $\bold M_{SP}^2$. The scalar--scalar
block $\bold M_{SS}^2$ and the pseudoscalar--pseudoscalar 
block $\bold M_{PP}^2$ can be found 
in Ref.~\cite{NuMass} for real parameters. 
They are valid also for the complex case after 
obvious substitutions for complex parameters.
\be{MSP}
{\bold M_{SP}^2}=
\left[\matrix{ \ds
0 & 
|B\mu| s_{\psi}&
-|\mu||\epsilon_1| s_{\rho_1}&
-|\mu||\epsilon_2|s_{\rho_2}& 
-|\mu||\epsilon_3|s_{\rho_3}
\cr 
\vb{18}
|B\mu| s_{\psi} & 
0 &
-|B_1\epsilon_1|s_{\sigma_1} &
-|B_2\epsilon_2|s_{\sigma_2}&
-|B_3\epsilon_3|s_{\sigma_3}
\cr
\vb{18}
|\mu||\epsilon_1| s_{\rho_1} & 
-|B_1\epsilon_1|s_{\sigma_1}&
0&
|\epsilon_1||\epsilon_2| s_{\lambda_{12}} &
|\epsilon_1||\epsilon_3| s_{\lambda_{13}}
\cr 
\vb{18}
|\mu||\epsilon_2| s_{\rho_2} &
-|B_2\epsilon_2|s_{\sigma_2}&
-|\epsilon_1||\epsilon_2| s_{\lambda_{12}}&
0 &
|\epsilon_2||\epsilon_3| s_{\lambda_{23}}
\cr 
\vb{18}
|\mu||\epsilon_3| s_{\rho_3}&
-|B_3\epsilon_3|s_{\sigma_3}&
-|\epsilon_1||\epsilon_3| s_{\lambda_{13}}&
-|\epsilon_2||\epsilon_3| s_{\lambda_{23}} & 
0
}\right],
\ee
where we have used the shorthand notation:
$s_{\psi}\equiv \sin(\theta+\phiB), s_{\rho_i}\equiv
\sin(\phim+\theta-\phiei-\eta_i),s_{\sigma_i}\equiv
\sin(\eta_i+\phiBi)$ and $s_{\lambda_{ij}}\equiv
\sin(\phiei+\eta_i-\varphi_{\epsilon_j}-\eta_j)$.

Eq.~\rf{Tadodd} together with Eq.~\rf{MSP}
shows that scalar--pseudoscalar Higgs transitions appear
already at tree--level,
contrary to the MSSM where this is possible only at one loop
level.
However, considering the typical size
of the \rp violating parameters, 
$|\epsilon_i/\mu|\lesssim O(10^{-3})$ \cite{NuMass}, such
a transition has to be expected to be rather small.

\subsection{Neutralino--neutrino mass matrix}

Here we focus on the $7\times 7$ Neutralino--neutrino
mass matrix. In the basis\\
${\Psi'_{0}}^T =(-i\lambda_1, -i \lambda_2,
\tilde{H}_d^1,\tilde{H}_u^2,
\nu_e,\nu_{\mu},\nu_{\tau})$, it reads

\be{nmm}
{\cal M}^0 =  \left(
                    \begin{array}{cc}
                    {\cal M}_{\chi^0} & m^T\\
                      m & 0 \\
                    \end{array}
              \right).
\ee
Here, the $3\times 4$  sub-matrix $m$ contains 
entries from the bilinear \rp 
parameters,
\be{bnmm}
m =   \left(
            \begin{array}{cccc}
     -\frac{1}{2}g' v_1 e^{-i\eta_1} & \frac{1}{2}g v_1 e^{-i\eta_1}
& 0 & |\epsilon_1|e^{i\phiee} \\
-\frac{1}{2}g' v_2 e^{-i\eta_2}& \frac{1}{2}g v_2 e^{-i\eta_2}
 & 0 & |\epsilon_2|e^{i\phiem} \\
        -\frac{1}{2}g' v_3 e^{-i\eta_3}& \frac{1}{2}g v_3 e^{-i\eta_3} & 
          0 & |\epsilon_3| e^{i\phiet}\\
                    \end{array}
              \right),
\ee
and the MSSM neutralino mass matrix is given by,

\be{MSSMnm}  
{\cal M}_{\chi^0} =  \left(
                        \begin{array}{cccc}
 |M_1| e^{i\phi_1} & 0   & -\frac{1}{2}g' v_d e^{-i\theta} &
\frac{1}{2} g' v_u  \\
 0   & |M_2| e^{i\phi_2} &  \frac{1}{2}g  v_d e^{-i\theta}&
 -\frac{1}{2} g  v_u  \\
  -\frac{1}{2} g' v_d e^{-i\theta}&  \frac{1}{2} g v_d e^{-i\theta}
& 0 & -|\mu|e^{i\phim}  \\
   \frac{1}{2} g' v_u & -\frac{1}{2} g v_u & 
-|\mu|e^{i\phim} & 0 \\
 \end{array}
                     \right).
\ee
The mass matrix ${\cal M}^0$ is diagonalized by
\beq
{{\cal  N}^0}^*{\cal M}^0{{\cal N}^0}^{-1}=
{\rm diag}(m_{\nu_i},m_{\chi^0_j}).
\label{chi0massmat}
\eeq
where $(i=1,{}_{\cdots},3)$ for the neutrinos, 
and $(j=1,{}_{\cdots},4)$ for
the neutralinos.
The method of perturbative diagonalization of ${\cal M}^0$ 
presented in \cite{SmallExp} holds also in the general case
of a complex symmetric mass matrix ${\cal M}^0$.

Only one neutrino acquires mass at tree--level and its mass is 
approximately given by
\beq
\label{mnutree}
m_{\nu_3} \simeq 
\Biggl|\frac{M_1 g^2 + M_2 {g'}^2}{4\, \det({\cal M}_{\chi^0})}\Biggr|
\,|{\vec \Lambda}|^2
\eeq
where,
\be{deflam}
{\Lambda_i} = |\epsilon_i| v_d \, e^{i(\phiei-\theta)}+ 
v_i|\mu|\, e^{i(\phim-\eta_i)}.
\ee
As a result, for a realistic description of the
neutrino spectrum one has to improve the calculation
to 1-loop order. 

A complete list of 1-loop contributions can be found, for example, 
in \cite{NuMass}. 
For our purposes it is sufficient to consider only the $\tilde b -b$ loop, 
which gives in a wide range of parameter space the most important 
contribution to the neutrino masses. The 1-loop corrections to 
the neutrino sector can be written as
\be{bloop}
(\Delta m_{\nu})^{\tilde b b}_{ij}\simeq -
\frac{3}{16\pi^2} m_b h_b^2 
\sin2\theta_{\tilde b} \, e^{i\varphi_{\tilde b}}
{{\cal N}^0_{i3}}^{\ast}\, {{\cal N}^0_{j3}}^{\ast}
\log\Biggl(\frac{m^2_{\tilde b_2}}{m^2_{\tilde b_1}}\Biggr ),
\ee
where $h_b=e^{-i\theta}\sqrt{2}m_b/v_d$,
$\theta_{\tilde b}$ is the
mixing angle and $\varphi_{\tilde b}$ is
the phase in the sbottom sector, respectively.
$m_{\tilde b_i}$ are the two sbottom masses.
The eigenvalues up to one loop are obtained 
by diagonalizing ${\rm diag}(m_{\nu_i},m_{\chi^0_j})+
(\Delta m_{\nu})^{\tilde b b}_{ij}$.
The corresponding mixing matrix is ${\cal N}^1$,
and the complete mixing matrix which
relates weak basis and eigenbasis, is
\be{fullN}
{\cal N}={\cal N}^1\cdot{\cal N}^0. 
\ee

The CP violating Dirac phase $\delta$ entering
the oscillation formula can be extracted by
using the following relation
\be{Diracd}
|\delta|=\sin^{-1}\Biggl(\Biggl|\frac{8\, 
{\Im}m({\cal N}_{25}{\cal N}^{\ast}_{26}
{\cal N}_{16}{\cal N}^{\ast}_{15})}
{\cos\theta_{13}\sin2\theta_{13}\sin2\theta_{12}
\sin2\theta_{23}}\Biggr|\Biggr),
\ee
where $\theta_{13}=\sin^{-1}(|{\cal N}_{35}|),
\theta_{12}=\tan^{-1}(|{\cal N}_{25}/{\cal N}_{15}|)$ and
$\theta_{23}=\tan^{-1}(|{\cal N}_{36}/{\cal N}_{37}|)$.
Note that the quantity in the denominator of Eq.~\rf{Diracd}
is one of six equivalent representations of the Jarlskog invariant 
in the neutrino sector.

\section{Neutrino data and CP violating phases}

\subsection{Analytical discussion}

Data from neutrino oscillation experiments require that the 
bilinear R-parity breaking parameters 
must obey certain conditions \cite{NuMass}. 
As briefly discussed in the introduction, atmospheric 
neutrino data requires $\Delta m^2_{Atm}$ to be of 
the order of $\Delta m^2_{Atm} \simeq {\cal O}(10^{-3})$ eV$^2$, 
while the solar neutrino problem can be solved for 
$\Delta m^2_{\odot} \sim$ (few) $10^{-5}$ eV$^2$ in the 
case of the LMA solution. \footnote{The LOW solution 
$\Delta m^2_{\odot} \sim 10^{-7}$ eV$^2$ and the quasi 
vacuum solution $\Delta m^2_{\odot} \sim 10^{-9}$ eV$^2$ 
can not be excluded at present, but are strongly disfavoured 
after the SNO neutral current measurement, see for example 
\cite{Bahcall:2002hv}.} 

In bilinear R-parity breaking the neutrino spectrum is hierarchical 
and therefore $m_{\nu_3}$ and $m_{\nu_2}$ are approximately given by 
$m_{\nu_3} \simeq \sqrt{\Delta m^2_{Atm}} \simeq 0.05$ eV 
and $m_{\nu_2} \simeq \sqrt{\Delta m^2_{\odot}}$. Since, as 
explained above, at tree level the model has only one non-zero 
mass eigenstate, it is straightforward to see that 
the ratio of $m_{\nu_2}/m_{\nu_3} \simeq 
\sqrt{\Delta m^2_{\odot}}/\sqrt{\Delta m^2_{Atm}}$ determines the 
relative importance of the one-loop mass with respect to the tree-level 
mass in our model. 

The hierarchy of $m_{\nu_2}/m_{\nu_3} \simeq 0.1$ (in 
case of the LMA solution) implies that the terms in Eq.~\rf{deflam} 
must almost cancel.
Thus, three pairs of phases have to obey the condition
\be{phasecorr}
\phiei + \eta_i - \theta -\phim  \pm \pi < O(10^{-2})~.
\ee
This observation reduces 
the effective number of free phases in the bilinear model 
from six to three, as we will now demonstrate.

The phases $\eta_i$ have to obey the tadpole equations
Eq.~\rf{Tadodd} for an arbitrary set of input parameters.
In the limit $\phiei + \eta_i - \theta -\phim \pm \pi= 0$, 
the tadpole equations Eq.~\rf{Tadodd} reduce to
\baq{redTadd}
v_u v_d |B\mu|\sin(\phiB+\theta)&=&0~,\nonumber \\
v_i v_u |B_i\epsilon_i|\sin(\phiBi+\eta_i)&=&0~,\qquad i=1,2,3
\eaq
where the equation for $\theta$ is just the well-known MSSM relation.
To find the correct minimum of the potential, we get
from Eq.~\rf{redTadd} ~$\theta=-\phiB$ and 
$\eta_i=\pm\pi-\phiBi$.
Eq.~\rf{phasecorr} now reads
$\phiei = \phiBi -\phiB +\phim$ modulo $2\pi$, which means
that the number of independent 
phases is reduced to three.
This leads immediately to the important result
that the scalar--pseudoscalar mixing vanishes.
This can be read of 
directly from Eq.~\rf{MSP} since every single 
term is zero in this limit.

Atmospheric neutrino measurements provide an additional constraint 
on the bilinear model, due to the large angle in 
$\nu_{\mu} \leftrightarrow \nu_{\tau}$ oscillations, which 
has to be 
$\tan^2\theta_{Atm} \simeq 
\frac{|\Lambda_2|^2}{|\Lambda_3|^2}\simeq 1$. 
For non-zero phases this 
condition could only be met if 
\be{phasecorr2}
\phiem + \eta_2 \simeq \pm (\phiet + \eta_3)~.
\ee
We will investigate below to which extent Eq.~\rf{phasecorr2} has 
to be satisfied numerically. 

\subsection{Numerical discussion}

Before we describe our numerical analysis we specify the parameters.
We remove two unphysical phases by applying the
Peccei--Quinn symmetry $U(1)_{PQ}$ and the $R$ symmetry $U(1)_R$.
We choose $\phiB=0$ and $\phim=\pi$, and set all other phases
of the MSSM part to zero.
As input phases we take $\phiei$ and $\eta_i$
randomly in the range $[-\pi,\pi]$, and
use the stationary conditions
Eq.~\rf{Tadodd} to
solve for the phases $\phiBi$.
In order to reduce the number of parameters,
the numerical calculations were performed
in a constrained version of the MSSM.
We have scanned the parameters in the
following ranges:
$M_2\in [0,1.2]$ TeV, $|\mu| \in [0,2.5]$ TeV,
$m_0\in [0,1.2]$ TeV, $A_0/m_0$ and $B_0/m_0\in
[-3,3]$ and $\tan\beta\in [2.5,10]$. 
All randomly generated points were subsequently
tested for consistency with the stationary
conditions Eq.~\rf{Tadeven} and \rf{Tadodd}.

\begin{figure}[H]
\setlength{\unitlength}{1mm}
\begin{center}
\begin{picture}(140,50)
\put(-40,-200)
{\mbox{\epsfig{figure=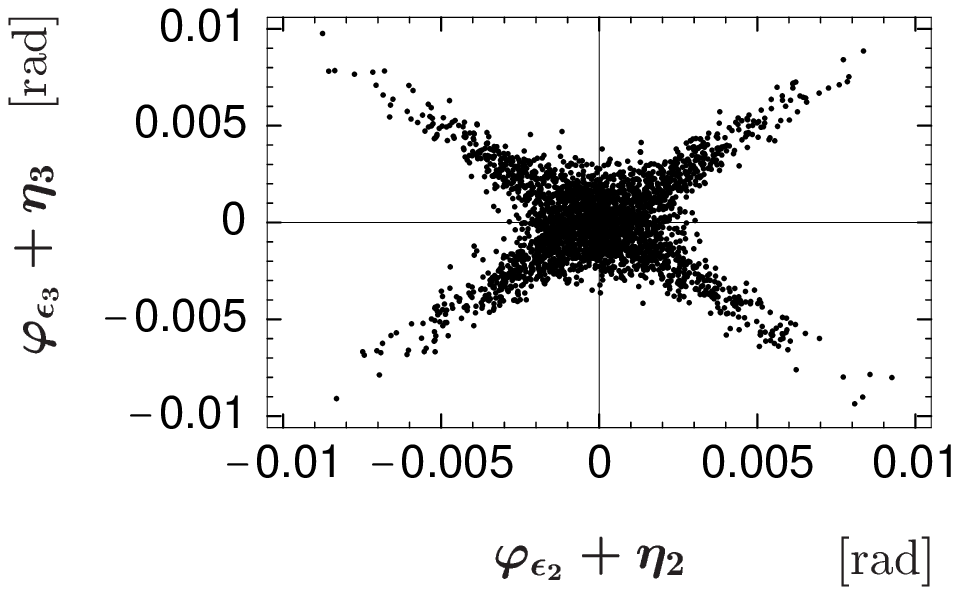,height=30.cm,width=20.cm}}}
\end{picture}
\end{center}
\caption{Phase correlation between 
$\phiem+\eta_2$ and $\phiet+\eta_3$ required by 
the constraints from atmospheric neutrino data.}
\label{sumd2d3}
\end{figure}

\begin{figure}[H]
\setlength{\unitlength}{1mm}
\begin{center}
\begin{picture}(140,50)
\put(-40,-200)
{\mbox{\epsfig{figure=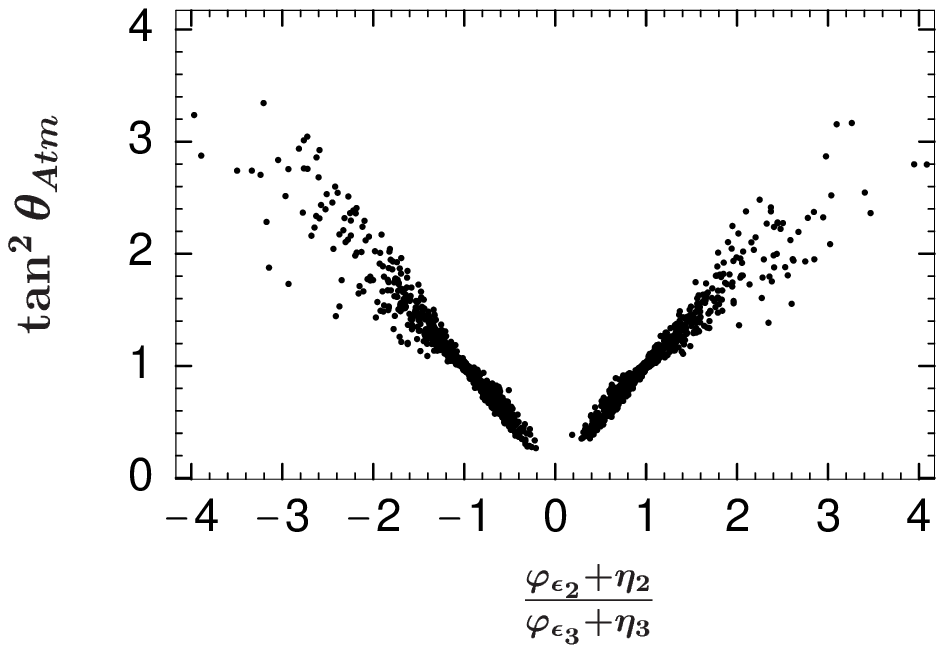,height=30.cm,width=20.cm}}}
\end{picture}
\end{center}
\vskip 0.5cm
\caption{Correlation between $(\phiem+\eta_2)/(\phiet+\eta_3)$
and $\tan^2\theta_{Atm}$.}
\label{corrd2d3}
\end{figure}

\begin{figure}[H]
\setlength{\unitlength}{1mm}
\begin{center}
\begin{picture}(140,50)
\put(-40,-200){\mbox
{\epsfig{figure=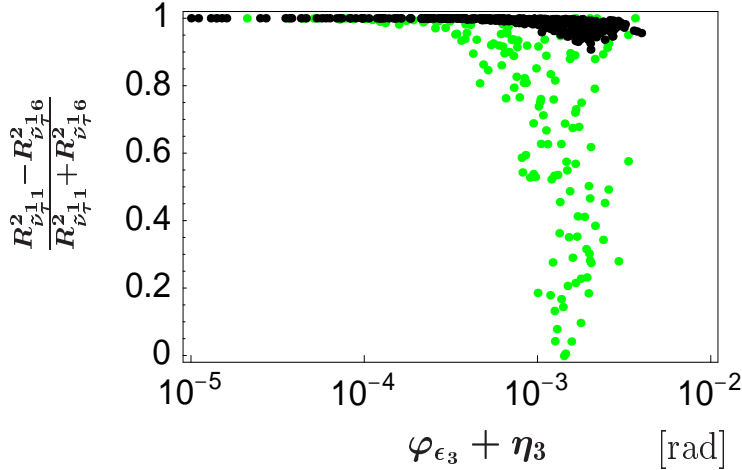,height=30.cm,width=20.cm}}}
\end{picture}
\end{center}
\caption{Scalar--pseudoscalar mixing in the nearly degenerate
$\tilde\nu^R-\tilde\nu^I$ system. For a discussion see text.}
\label{cpmixing}
\end{figure}

Next we consider to which
extent the correlations \rf{phasecorr} have to be obeyed.
With our choice of the phases these correlations read: 
$\phiei+\eta_i=0$ modulo $2\pi$.
We study the allowed region of the two sums of phases
$\phiem+\eta_2$ and $\phiet+\eta_3$.
From Fig.~\ref{sumd2d3} we can see 
$|\varphi_{\epsilon_{2,3}}+\eta_{2,3}|\lesssim 5\times 10^{-3}$
for most of the sums $\varphi_{\epsilon_{2,3}}+\eta_{2,3}$. 
If $\phiem+\eta_2\simeq \pm(\phiet+\eta_3)$, then
$|\varphi_{\epsilon_{2,3}}+\eta_{2,3}|\lesssim 10^{-2}$
is possible. This feature can be traced to the constraint
$\tan^2\theta_{Atm} \simeq 1$, see Eq.~\rf{phasecorr2}.
Fig.~\ref{corrd2d3} confirms the additional phase correlation given
in Eq.~\rf{phasecorr2}, where we have plotted
$\tan^2\theta_{Atm}$ versus $(\phiem+\eta_2)/(\phiet+\eta_3)$ for points 
in which $\varphi_{{\epsilon}_{2,3}}+\eta_{2,3} \ge 3\times 10^{-3}$. 
As has been pointed out previously, ${\bold M_{SP}}$=0
in the limit $\phiei+\eta_i=0$. 
Even if this condition is not exactly fulfilled
scalar--pseudoscalar Higgs mixing can be neglected in general
provided that $|\phiei+\eta_i|\lesssim  10^{-2}$.
In the sneutrino sector the situation may be
sligthly different.
Recall that in the MSSM the sneutrino is a complex
scalar field. \rp splits $\Re e(\tilde\nu)$ and $\Im m(\tilde\nu)$
by a small amount in each generation.
Let us denote the associated mass eigenstates as ${\tilde\nu}^{1,2}$.
Since ${\tilde\nu}^{1,2}$ are nearly degenerate
large scalar--pseudoscalar mixing
is possible in this sector
also for $|\phiei+\eta_i|\lesssim  10^{-2}$.
To demonstrate this feature, let us consider the third
generation ${\tilde\nu_\tau}^{1,2}$.
As an example we consider the
\rp coupling $C({\tilde\nu_{\tau}}^1 b b)$,
where ${\tilde\nu_{\tau}}^1$ does not contain
imaginary parts  
in the limit $\phiei+\eta_i=0$. 
The interaction is described by the Lagrangian
\be{riue}
{\mathcal L}={\tilde\nu_{\tau}}^1 \bar b \ [C_{{\tilde\nu_{\tau}}^1 b b}P_L+
C^{\ast}_{{\tilde\nu_{\tau}}^1 b b}P_R]\ b,
\ee
where $C_{{\tilde\nu_{\tau}}^1 b b}\equiv
-\frac{h_b}{\sqrt{2}}(R_{{\tilde\nu_{\tau}}^1 1}+
i R_{{\tilde\nu_{\tau}}^1 6})$ and $R$ is defined
in Eq.~\rf{Rotmat}. 
CP invariance means that $R_{{\tilde\nu_{\tau}}^1 6}=0$.
The extent to which this limit is saturated can be 
seen in Fig.~\ref{cpmixing} where we
plot $(R^2_{{\tilde\nu_{\tau}}^1 1}-R^2_{{\tilde\nu_{\tau}}^1 6})/
(R^2_{{\tilde\nu_{\tau}}^1 1}+R^2_{{\tilde\nu_{\tau}}^1 6})$ as a function
of $\phiet+\eta_3$.
For the light (dark) points the mass spliting is 
$10^{-3}$ eV $\leq |m_{{\tilde\nu^1}_{\tau}}-m_{{\tilde\nu^2}_{\tau}}|
\leq 10^{-2}$ eV (0.1 eV $\leq$ 
$|m_{{\tilde\nu^1}_{\tau}}-m_{{\tilde\nu^2}_{\tau}}|$).
As can be seen, scalar--pseudoscalar mixing 
in the sneutrino sector
vanishes only for very small $\phiei+\eta_i$ in this
sector.
\section{Observables}
In this section we discuss possible
CP--odd observables arising due to the new
phases in bilinear R-parity violating couplings. 
Recall that these \rp couplings are typically two to three
orders of magnitude smaller than the $R$ parity conserving couplings
if neutrino data are to be explained by R-parity violation.
This already implies that CP violating effects induced by R-parity
violating parameters can at most be seen in LSP decays, because in all other
cases either the R-parity violating branching ratios are very small or
the loop contributions due to the \rp couplings to R-parity conserving
decay modes are tiny. 

In the following we will discuss various possibilities for the LSP:
sfermions, neutralino and chargino. Here we will focus on observables
arising in two--body decays of the LSP, which are either rate asymmetries
or helicity asymmetries. Note that the assumption of two--body decays
is well motivated by the fact that LEP has not found any supersymmetric 
particle which gives a lower bound on SUSY particle masses 
of order 100 GeV.

Let us first discuss a necessary condition whether a CP asymmetry is
observable or not before considering the different LSP classes.
The relevant quantity to decide whether an asymmetry is observable
(at $1\sigma$), is given by  
$(A^2_{\rm CP}\times B)^{-1}$, where 
$A_{\rm CP}$ is the CP asymmetry and $B$
is the branching ratio of the decay considered.
If we assume rather optimistically that order $10^6$ LSPs 
can be produced at a future collider experiment, this requires 
that $(A^2_{\rm CP}\times B)^{-1}\lesssim O(10^6)$.
 
We want to recapitulate here, that in order to
construct a rate asymmetry one needs final--state
interactions, otherwise partial decay rates are equal 
due to CPT invariance even
if CP is violated.
The rate asymmetry is then built through an
interference of tree--level amplitudes
and one--loop amplitudes where a pair
of intermediate particles in the loop is 
on--shell.
Subsequently we will discuss various one--loop diagrams giving
rise to CP asymmetries.

\subsection{Sfermion LSP}

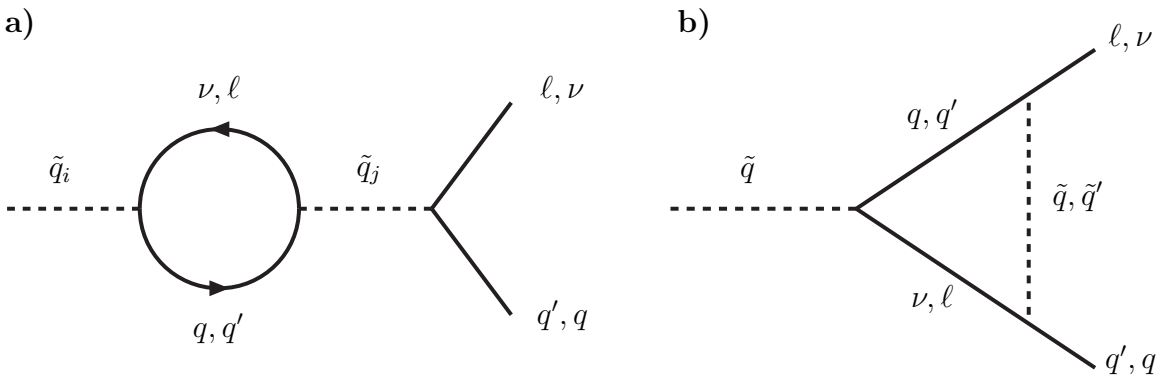
\begin{figure}[ht]
\begin{picture}(90,130)(10,0) 
\SetWidth{1.5} 
\DashLine(10,60)(60,60){3}
\ArrowArc(90,60)(30,180,0) 
\ArrowArc(90,60)(30,0,180)
\DashLine(120,60)(170,60){3}
\Line(170,60)(200,100)
\Line(170,60)(200,20) 
\Text(30,75)[]{{\large $\tilde q_i$}}
\Text(147,75)[]{{\large $\tilde q_j$}}
\Text(220,105)[]{{\large $\ell,\nu$}}
\Text(220,20)[]{{\large $q',q$}} 
\Text(90,105)[]{{\large $\nu,\ell$}} 
\Text(90,15)[]{{\large $q,q'$}} 
\Text(15,130)[]{\large  \bf{a)}}
\DashLine(260,60)(330,60){3}
\Line(330,60)(420,0)
\Line(330,60)(420,120)
\DashLine(395,20)(395,100){3}
\Text(290,75)[]{{\large $\tilde q$}}
\Text(435,125)[]{{\large $\ell,\nu$}}
\Text(435,3)[]{{\large $q',q$}}
\Text(360,96)[]{{\large $q,q'$}} 
\Text(360,26)[]{{\large $\nu, \ell$}} 
\Text(415,65)[]{{\large $\tilde q, \tilde q'$}}
\Text(270,130)[]{\large  \bf{b)}}
\end{picture}
\caption{a) Self energy diagrams and b) vertex diagrams contributing 
to CP asymmetries of squark decays.\label{fig:squarklsp}}
\end{figure}

Consider first the case were a squark is the LSP because here the 
discussion is rather simple. The possible final states are either 
$q' \, \ell_i$ or $q \, \nu_j$.
The relevant diagrams contributing to possible CP
asymmetries are shown in Fig.~\ref{fig:squarklsp}: a) self energy 
diagrams and b) vertex diagrams. It is obvious from this figure that 
in both cases all
three couplings involved violate R--parity and, thus, the corresponding
diagram gives only a tiny contribution leading to a negligible asymmetry.

%
%
%
%
%
%
\begin{figure}[H]
\begin{picture}(120,290)(10,0) 
\SetWidth{1.5} 
\DashLine(10,60)(60,60){3}
\ArrowArc(90,60)(30,180,0) 
\ArrowArc(90,60)(30,0,180)
\DashLine(120,60)(170,60){3}
\Line(170,60)(200,100)
\Line(170,60)(200,20) 
\Text(30,75)[]{{\large $\tilde \ell_i$}}
\Text(147,75)[]{{\large $H^\pm$}}
\Text(212,105)[]{{\large $\ell$}}
\Text(212,20)[]{{\large $\nu$}} 
\Text(90,105)[]{{\large $\ell,\nu, q$}} 
\Text(90,15)[]{{\large $\nu, \ell, q'$}} 
\Text(10,130)[]{\large  \bf{c)}}
\DashLine(260,60)(330,60){3}
\Line(330,60)(420,0)
\Line(330,60)(420,120)
\DashLine(395,18)(395,102){3}
\Text(290,75)[]{{\large $\tilde \ell$}}
\Text(435,125)[]{{\large $\nu$}}
\Text(435,3)[]{{\large $\ell$}}
\Text(360,96)[]{{\large $\ell$}} 
\Text(360,26)[]{{\large $\nu$}} 
\Text(415,65)[]{{\large $H^\pm$}}
\Text(270,130)[]{\large  \bf{d)}}
%

\DashLine(10,220)(60,220){3}
\ArrowArc(90,220)(30,180,0) 
\ArrowArc(90,220)(30,0,180)
\Photon(120,220)(170,220){3}{6.5}
\Line(170,220)(200,260)
\Line(170,220)(200,180) 
\Text(30,235)[]{{\large $\tilde l_i$}}
\Text(147,235)[]{{\large $W$}}
\Text(212,265)[]{{\large $\ell$}}
\Text(212,180)[]{{\large $\nu$}} 
\Text(90,265)[]{{\large $\ell,\nu, q$}} 
\Text(90,175)[]{{\large $\nu, \ell, q'$}} 
\Text(10,290)[]{\large  \bf{a)}}
\DashLine(260,220)(330,220){3}
\Line(330,220)(420,160)
\Line(330,220)(420,280)
\Photon(395,178)(395,262){3}{11}
\Text(290,235)[]{{\large $\tilde \ell$}}
\Text(435,285)[]{{\large $\nu$}}
\Text(435,163)[]{{\large $\ell$}}
\Text(360,256)[]{{\large $\ell$}} 
\Text(360,186)[]{{\large $\nu$}} 
\Text(415,225)[]{{\large $W$}}
\Text(270,290)[]{\large \bf{b)}}
\end{picture}
\caption{a) and c) Self energy diagrams and b) and d)
 vertex diagrams contributing 
to CP asymmetries of slepton decays.\label{fig:sleptonlsp}}
\end{figure}
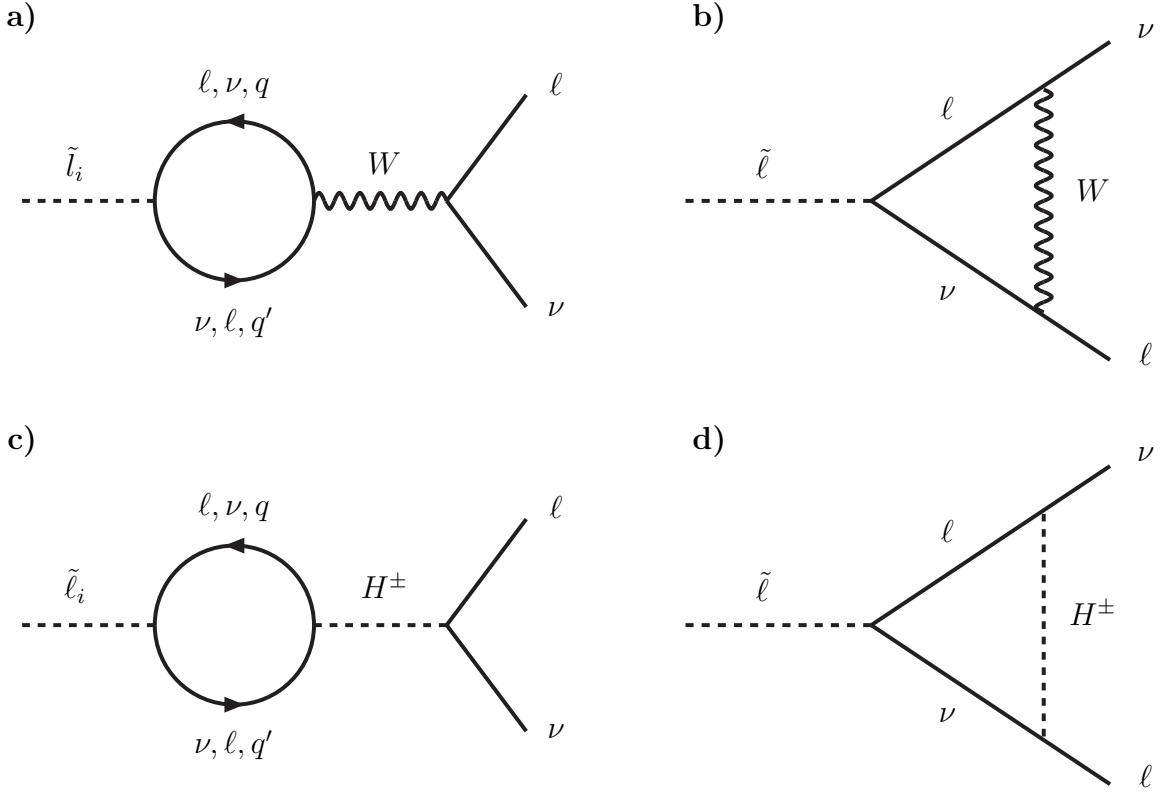
For a charged slepton ($\tilde \ell$) LSP the possible final states 
are $q q'$ and $\ell\nu$. 
We consider the following rate asymmetry, 
\be{slepasym}
{\mathcal A}_{\rm CP}=\frac{\Gamma(\tilde\ell\to \ell^-\nu)- 
\Gamma(\bar{\tilde\ell}\to \ell^+\nu)}{\Gamma(\tilde\ell\to \ell^-\nu)+
\Gamma(\bar{\tilde\ell}\to \ell^+\nu)}~.
\ee
In this case diagrams similar 
to those shown in Fig. \ref{fig:squarklsp} with appropriate 
replacements contribute. For the same reasoning as given above for the 
squark case their contribution is small. In addition to these diagrams 
one finds diagrams involving a $W$--boson and the corresponding
ones where the $W$--boson is replaced by  a charged Higgs boson 
as shown in Fig.~\ref{fig:sleptonlsp}.
As can be easily shown the diagrams involving a $W$--boson are suppressed
due to the small fermion masses involved.
In case of a self--energy
contribution with a top quark and a bottom quark 
in the loop, Fig.~\ref{fig:sleptonlsp}a, the CP asymmetry 
is suppressed by the factor $\frac{m_t m_{\ell}}{m_W^2}$. 
For the charged Higgs boson the situation is a little bit more subtle, 
because in the corresponding diagram of Fig.~\ref{fig:sleptonlsp}c the 
contribution with top/bottom quarks can be enhanced
for large $\tan\beta$.
In this case the CP asymmetry is given by
\be{asychHex}
{\cal A}_{\rm CP}\sim \frac{3}{16\pi}\frac{\Im m\{
h_{\ell} \sin\beta 
(g^L_1 h_b \sin\beta+g^R_1 h_t \cos\beta) g^{\ast}_4\}}{|g_4|^2}~,
\ee
where $g^{L,R}_1\ (g_4)$ are the \rp and CP violating left and right
(left) coupling of the $\tilde\ell$ to $t b$ ($\nu \ell$).
The relevant Lagrangian is given by
\be{Lagrangeslep}
{\mathcal L}=\tilde\ell \ \bar b (g_1^{L} P_L+g_1^{R} P_R)\ t+
g_4 \tilde\ell \ \bar\ell P_L\ \nu + {\rm h.c.}
\ee
where $P_{L,R}=1/2(1\mp\gamma_5)$.
The couplings can be found in \cite{NeuLSP}.
In principle large $\tan\beta$ may lead to a large CP asymmetry if 
$\ell=\tau$.
However, for large values of $\tan\beta$ the 1-loop contributions to 
the neutrino masses tend to be too large.
For this reason we have not found 
any points with $(A^2_{\rm CP}\times B)^{-1}\lesssim O(10^6)$,
satisfying at the same time the constraints from neutrino physics.

Let us now turn to the case where the sneutrino is the LSP. It can be 
shown on general grounds that the mass
splitting between $\Re e(\tilde \nu)$ and $\Im m(\tilde \nu)$ is very small
if the Majorana mass of the corresponding neutrino is tiny 
\cite{martinsnu}. Taking the known neutrino data into account this implies
that in our case the CP-even/CP-odd mass splitting between 
$\Re e(\tilde \nu)$ and $\Im m(\tilde \nu)$ is typically of order eV and 
thus negligibly small.
In the case of CP violation the mass eigenstates
are a superposition of
${\tilde\nu_{\ell}}^R$ and ${\tilde\nu_{\ell}}^I$.
As a possible CP sensitive observable we consider
the following helicity asymmetry,
\be{helicityasym}
{\mathcal A}_{\rm hel}=
\frac{\sum_i\lbrack\Gamma(\tilde\nu^i_{\ell}\to \tau^+_L \tau^-_L)-
\Gamma(\tilde\nu^i_{\ell}\to \tau^+_R \tau^-_R)\rbrack}
{\sum_i\lbrack\Gamma(\tilde\nu^i_{\ell}\to \tau^+_L \tau^-_L)+
\Gamma(\tilde\nu^i_{\ell}\to \tau^+_R \tau^-_R)\rbrack},
\ee
where $\ell=\tau,\mu,e$, and we sum over the two (nearly) degenerate 
sneutrino states, because they can not be resolved experimentally. 

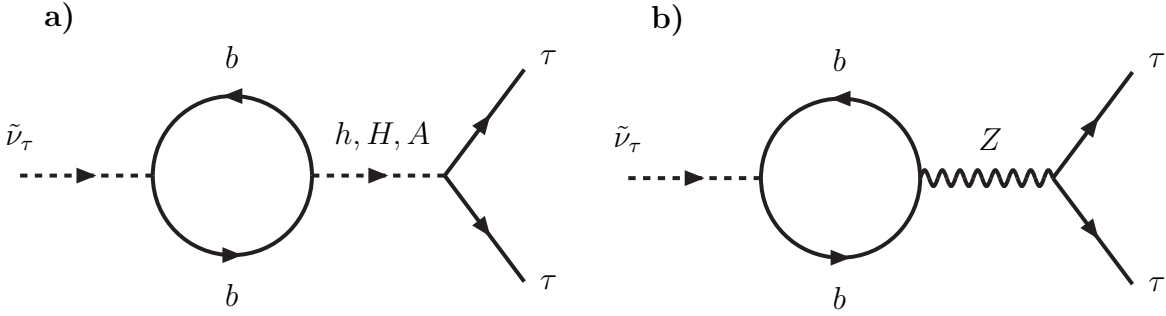
\begin{figure}[H]
\begin{picture}(100,130)(10,0) 
\SetWidth{1.5} 
\DashArrowLine(10,60)(60,60){3}
\ArrowArc(90,60)(30,180,0) 
\ArrowArc(90,60)(30,0,180)
\DashArrowLine(120,60)(170,60){3}
\ArrowLine(170,60)(200,100)
\ArrowLine(170,60)(200,20) 
\Text(10,75)[]{{\large $\tilde\nu_{\tau}$}}
\Text(147,75)[]{{\large $h,H,A$}}
\Text(210,105)[]{{\large $\tau$}}
\Text(210,20)[]{{\large $\tau$}} 
\Text(90,105)[]{{\large $b$}} 
\Text(90,15)[]{{\large $b$}} 
\Text(25,120)[]{\large  \bf{a)}}
\end{picture}

\begin{picture}
(100,130)(-220,-130) 
\SetWidth{1.5} 
\DashArrowLine(10,60)(60,60){3}
\ArrowArc(90,60)(30,180,0) 
\ArrowArc(90,60)(30,0,180)
\Photon(120,60)(170,60){3}{7.5}
\ArrowLine(170,60)(200,100)
\ArrowLine(170,60)(200,20) 
\Text(10,75)[]{{\large $\tilde\nu_{\tau}$}}
\Text(147,75)[]{{\large $Z$}}
\Text(210,105)[]{{\large $\tau$}}
\Text(210,20)[]{{\large $\tau$}} 
\Text(90,105)[]{{\large $b$}} 
\Text(90,15)[]{{\large $b$}}
\Text(25,120)[]{\large  \bf{b)}} 
\end{picture}

\vskip-4cm
\caption{Dominant Feynman diagrams for the absorptive 
part of the amplitude $\tilde\nu^i\to\tau\tau$.}
\label{sneutrinolsp}
\end{figure}

\begin{figure}[H]
\setlength{\unitlength}{1mm}
\begin{center}
\begin{picture}(140,50)
\put(-40,-200)
{\mbox{\epsfig{figure=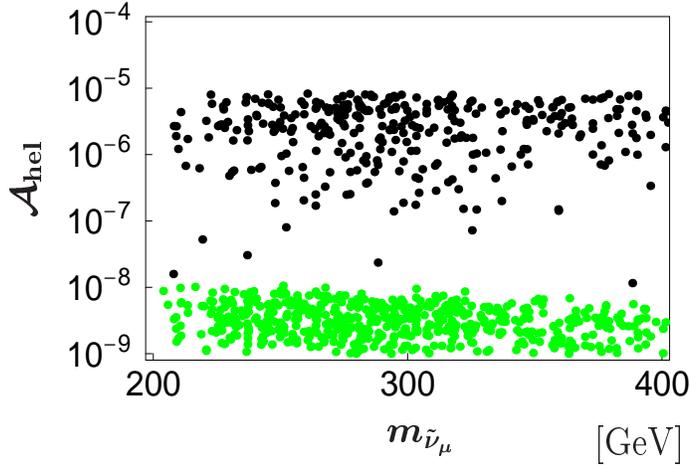,height=30.cm,width=20.cm}}}
\end{picture}
\end{center}
\caption{${\mathcal A}_{\rm hel}$ 
as a function 
of $m_{\tilde\nu_{\mu}}$ where the dark (light) points
represent the result without (with) sum of the
contribution of the two degenerate ${\tilde\nu^i_{\mu}}$.
\label{cancelasym} }
\end{figure}

The same classes of diagrams appear as in the case of the charged slepton
decays discussed previously. The dominant diagrams are now the 
self--energy diagrams with a bottom quark in the loop, see 
Fig.~\ref{sneutrinolsp}.
The contributions where a $Z$ is exchanged in the s--channel
(Fig.~\ref{sneutrinolsp}b) 
is suppressed by a factor $\frac{m_{\tau} m_b}{m^2_Z}$.
The contribution in Fig.~\ref{sneutrinolsp}a 
can be substantial for large
values of $\tan\beta$. However, as was mentioned,
this in turn tends to drive the 
1-loop contribution to the neutrino masses to be too large.
Moreover,
there is a cancellation between the contributions
of the two degenerate sneutrinos for $\tilde\nu^i_{\mu, e}$.
This cancellation can be seen from Fig.~\ref{cancelasym}
for $\tilde\nu^i_{\mu}$ where
the dark points represent the result for $i=1$
in Eq.~\rf{helicityasym}, whereas for the light points
the two contributions are summed up.
For $\tilde\nu_e$ the result is the same.
For $\tilde\nu^i_{\tau}$ we find that
$\sum_i B({\tilde\nu_{\tau}}^i\to\tau^+\tau^-)\sim O(10^{-5}-10^{-4})$.
With the expected number of events, it is practically 
impossible to observe such a helicity asymmetry.  
%
%
%
\subsection{Neutralino LSP}

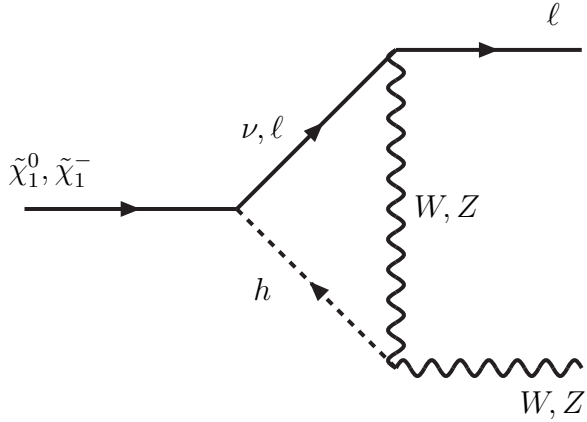
\begin{figure}[t]
\begin{center}
\begin{picture}(250,185)(0,0)
\SetWidth{1.5} 
\ArrowLine(0,60)(80,60)
\ArrowLine(80,60)(140,120)
\DashArrowLine(140,0)(80,60){3}
\Photon(140,0)(140,120){3}{10.5}
\Photon(140,0)(210,0){3}{6.5}
\ArrowLine(140,120)(210,120)
\Text(90,90)[]{{\large $\nu,\ell$}}
\Text(90,30)[]{{\large $h$}} 
\Text(10,75)[]{{\large $\tilde\chi^0_1,\tilde\chi^-_1$}}
\Text(200,-14)[]{{\large $W,Z$}}
\Text(160,60)[]{{\large $W,Z$}}
\Text(200,134)[]{{\large $\ell$}}
\end{picture}
\vskip 1cm
\caption{Dominant Feynman diagram for the absorptive 
part of the amplitude $\tilde\chi^0_1\to\ell\ W$ and
$\tilde\chi^-_1\to\ell\ Z$.\label{vertexgraph}}
\end{center}
\end{figure}

If the $\tilde\chi^0_1$ is the LSP the decay modes 
may be $\tilde\chi^0_1\to \{W \ell,Z \nu,h \nu\}$.
As an example we consider rate asymmetries such as 
\be{rateasym}
{\cal A}_{\rm CP}=
\frac{\Gamma(\tilde\chi^0_1\to W^+\ell^-)-
\Gamma(\tilde\chi^0_1\to W^-\ell^+)}
{\Gamma(\tilde\chi^0_1\to W^+\ell^-)+
\Gamma(\tilde\chi^0_1\to W^-\ell^+)}\ .
\ee
The relevant part of the Lagrangian reads
\be{LagrangeNeu}
{\cal L}=\bar{\nu} O_R^{\nu\chi h} P_R \ \tilde\chi^0_1 h
+\bar{\ell}\gamma^{\mu}
(O_R^{\ell\chi {\rm w}} P_R+
O_L^{\ell\chi {\rm w}} P_L) \tilde\chi^0_1 W^-_{\mu}
+\bar{\nu}\gamma^{\mu}
O_L^{\nu\chi {\rm z}} P_L\ \tilde\chi^0_1 Z_{\mu}+{\rm h.c.}
\ee
The relevant couplings in Eq.~\rf{LagrangeNeu} can be found in \cite{NeuLSP}. 
Typically these couplings obey:
$|O_R^{\ell\chi {\rm w}}|\sim |O_R^{\nu\chi h}|>>
|O_L^{\ell\chi {\rm w}}|,|O_L^{\nu\chi {\rm z}}|$. 
This implies that final state interactions between
$W\ell$ and $Z \nu$ will not contribute
significantly to the CP asymmetry in Eq.~\rf{rateasym}.
As the next possibility we consider a final--state interaction between 
$W \ell$ and $h \nu$, see Fig.~\ref{vertexgraph}.
The contribution where the $W^+$ is
replaced by the charged Higgs $H^+$, is 
suppressed due to the Yukawa coupling and
therefore not relevant.
Setting small lepton masses to zero,
a straightforward calculation gives an asymmetry
proportional to 
$\Im m(O_R^{\nu\chi h} {O_L^{\ell\chi {\rm w}}}^{\ast})$ (see Appendix).
Since only the left coupling $O_L^{\ell\chi {\rm w}}$
(and not the much larger right coupling
$O_R^{\ell\chi {\rm w}}$) appears in the denominator
of Eq.~\rf{rateasym} the resulting 
$(A^2_{\rm CP}\times B)^{-1}$ is always above $10^6$.
%
%
\vskip-2cm

\vskip 2.5cm

\subsection{Chargino LSP}

\begin{figure}[t]
\setlength{\unitlength}{1mm}
\begin{center}
\begin{picture}(140,50)
\put(-40,-200){\mbox
{\epsfig{figure=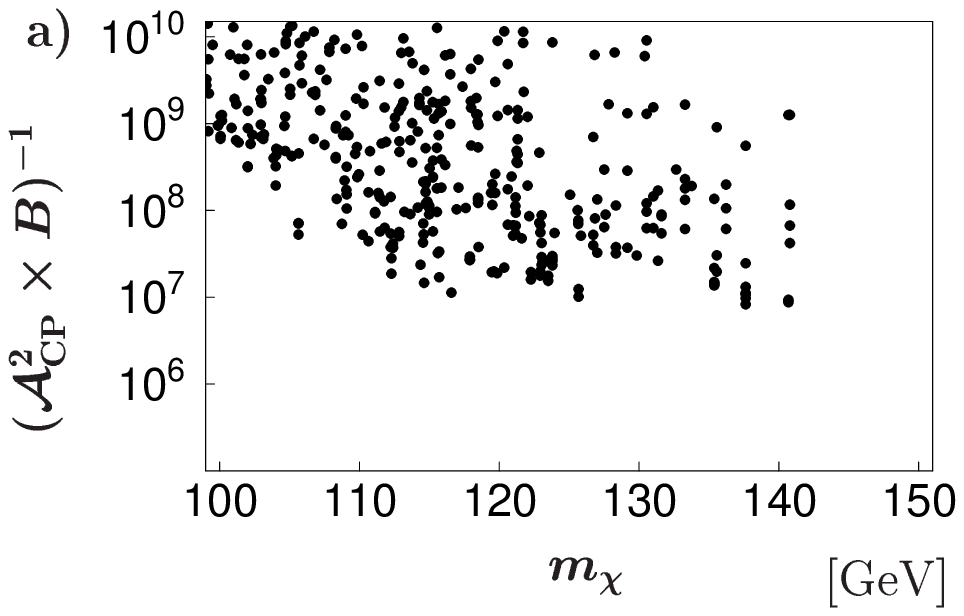,height=30.cm,width=20.cm}}}
\put(-40,-270){\mbox
{\epsfig{figure=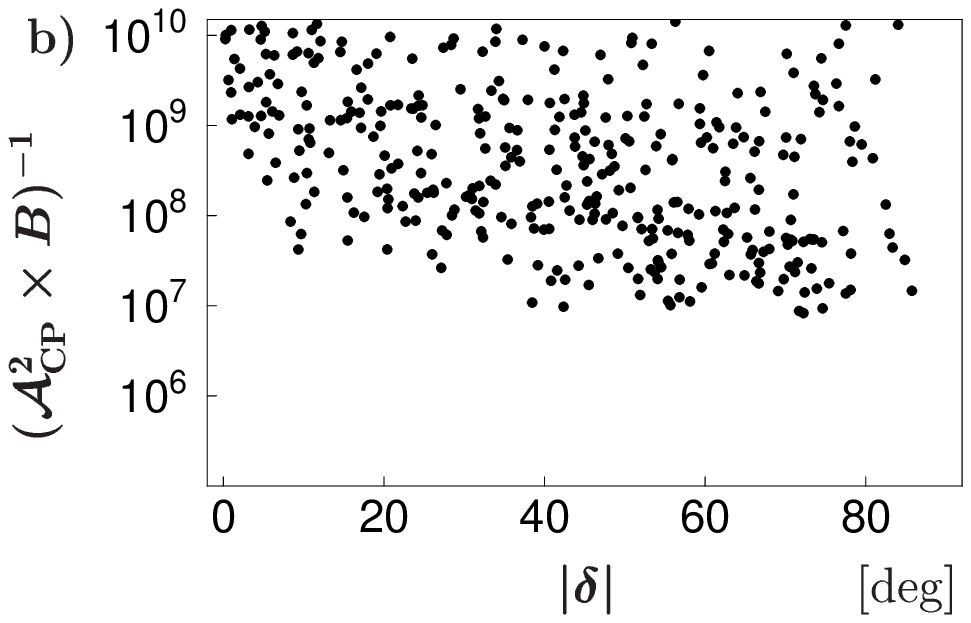,height=30.cm,width=20.cm}}}
\end{picture}
\end{center}
\vskip7.5cm
\caption{$({\mathcal A}^2_{\rm CP}\times B)^{-1}$ 
for $\chi^{\pm}\to Z e^{\pm}$ as a function of a) $m_{\chi}$
and b) $|\delta|$, respectively.}
\label{asym}
\end{figure}

If the $\chi^-_1$ is the LSP the possible final states 
are $Z \ell, W \nu$ and $h \, \ell$.
We consider the CP asymmetry: 
\be{CPasych}
{\cal A}_{\rm CP}=
\frac{\Gamma(\tilde\chi^-_1\to Z \ell^-)-\Gamma(\tilde\chi^+_1\to Z\ell^+)}
{\Gamma(\tilde\chi^-_1\to Z\ell^-)+\Gamma(\tilde\chi^+_1\to Z\ell^+)}
\ee
The relevant part of the Lagrangian reads
\be{Lagrange}
{\cal L}=\bar{\ell}(O_L^{\ell\chi h} P_L+
O_R^{\ell\chi h} P_R)\tilde\chi^-_1 h
+\bar{\nu}\gamma^{\mu}
O_L^{\nu\chi {\rm w}} P_L \tilde\chi^-_1 W_{\mu}
+\bar{\ell}\gamma^{\mu}(O_L^{\ell\chi {\rm z}} P_L+
O_R^{\ell\chi {\rm z}} P_R)\tilde\chi^-_1 Z_{\mu}+{\rm h.c.}
\ee
The full form of the left and right couplings can
be found in \cite{NuMass,NeuLSP}.
Scanning the parameters over the parameter ranges as described
in the previous section we find that the left and right couplings in 
Eq.~\rf{Lagrange} typically are:
$|O_L^{\nu\chi {\rm w}}|,|O_L^{\ell\chi {\rm h}}|,
|O_R^{\ell\chi {\rm z}}|$ $ << |O_R^{\ell\chi  h}|
\sim |O_L^{\ell\chi {\rm z}}|$.
Therefore, similar to the neutralino we will consider
only a final--state interaction between
$Z \ell$ and $h \ell$.
The dominant Feynman diagram is displayed in 
Fig.~\ref{vertexgraph}, while 
the calculation is given in the Appendix.
The result of a numerical scan is shown Fig.~\ref{asym}a and b where we plot
$({\mathcal A}_{\rm CP}^2\times B)^{-1}$ as a function of
$m_{\chi}$ and of the modulus of the 
Dirac phase $|\delta|$ defined in Eq.~\rf{Diracd}, respectively.
We show $({\mathcal A}^2_{\rm CP}\times B)^{-1}$
for $\ell=e$; for $\ell=\mu, \tau$
the result is similar.
As we can see in Fig.~\ref{asym}a and b, 
$({\mathcal A}^2_{\rm CP}\times B)^{-1}$ is 
always above $10^6$.
\section{Conclusion}

Supersymmetric models with bilinear R-parity breaking contain 
six new phases compared to the MSSM. These phases are currently 
only constrained by neutrino data. We have found that neutrino 
physics requires that the phases in the \rp sector have to fulfill 
the relation $\phiei \simeq  \phiBi$, but are otherwise not 
necessarily small.
This in turn implies that scalar--pseudoscalar
mixing is vanishingly small even though the phases may be maximal
CP violating.
The only exception is if a CP--even and a CP--odd state
are nearly degenerate, in which case the mixing between
these two states can be large.

We have discussed CP asymmetries in the decays of various possible 
LSPs and concluded that all of them are 
unmeasurably small. Although we have considered 
only one class of CP violating observables, namely rate asymmetries,
we conjecture that also other CP odd observables are to small to 
be measurable.

On the other hand, the neutrino oscillation Dirac phase $\delta$ 
is not necessarily small if bilinear \rp parameters are complex. 
This implies that possibly the only way to determine whether the 
\rp parameters are complex is to measure CP violation in 
neutrino oscillations themselves.

\vskip10mm
\acknowledgments
We thank A.~Bartl and J.~W.~F.~Valle for useful discussions.
This work was supported by Spanish grant BFM2002-00345, 
by the European Commission RTN networks HPRN-CT-2000-00148 and
HPRN-CT-2000-00149,
by the 
`Fonds zur F\"orderung der wissenschaftlichen Forschung'
of Austria FWF, Project No. P13139-PHY,
by Acciones Integradas Hispano--Austriaca
and by  
the European Science Foundation network grant N.~86. T.K. 
acknowledges financial support from 
the European Commission Research Training Site contract HPMT-2000-00124;
M.H. is supported by a MCyT Ram\'on y Cajal contract. 
 W.~P.~is supported 
by the 'Erwin Schr\"odinger fellowship' No.~J2095 of the `Fonds zur
F\"orderung der wissenschaftlichen Forschung' of Austria FWF and
partly by the Swiss `Nationalfonds'.

\section{Appendix}
Here we outline the calculation of the rate asymmetry
for neutralino and chargino decay.
The calculation is carried out in the 
Feynman--'t Hooft gauge.
In the limit where $m_{\ell}=0$
and $|O_R^{\ell\chi h}| >> |O_L^{\ell\chi h}|$
the dominant one--loop amplitude, corresponding
to the Feynman diagram in Fig.~\ref{vertexgraph}, 
has the same generic structure
for chargino and neutralino decays.
For the decay $\chi\to \ell^- V$, with
a negative charged lepton in the final state,
the amplitude reads:
\be{oneloop}
{\mathcal M}^{(1)}=\frac{i}{(4\pi)^2}\bar u(p_{\ell})
\lbrack 2 p_{\ell\rho} (C_0+C_1+C_2)P_R-
m_{\chi}C_2 \gamma_{\rho}P_L\rbrack u(p_{\chi})
{\varepsilon^\rho}^{\ast}(-p_V) g^L_1 g_0 \frac{2 m^2_V}{v}~,
\ee
$C_i$ being the Passarino--Veltman functions,
$g_0\equiv O_R^{\ell\chi h}
\ \lbrack O_R^{\nu\chi h}\rbrack$
and
$g^L_1\equiv -\frac{g}{\cos\theta_W}(-\frac{1}{2}+\sin^2\theta_W) 
\lbrack -\frac{g}{\sqrt 2}\
{\mathcal N}^{\ast}_{i(\ell+4)}\rbrack$
for chargino [neutralino] decay.
The subscript $V$ stands for the appropriate
vector boson and $\chi$ for the appropriate
SUSY particle, respectively.
The tree--level amplitude is given by:
\be{treelevel}
{\mathcal M}^{(0)}=i \bar u(p_{\ell})\gamma_{\rho}
\lbrack g^L_3 P_L + g^R_3 P_R\rbrack u(p_{\chi})
{\varepsilon^\rho}^{\ast}(-p_V)~,
\ee
where $g^L_3\equiv O_L^{\ell\chi {\rm z}}
\ \lbrack O_L^{\ell\chi {\rm w}}\rbrack$ and
$g^R_3\equiv O_R^{\ell\chi {\rm z}}
\ \lbrack O_R^{\ell\chi {\rm w}}\rbrack$
for chargino [neutralino] decay.
The CP asymmetry is now
\baq{CPasym}
{\cal A}_{\rm CP}&=&
\frac{\Gamma(\chi\to V \ell^-)-\Gamma(\bar{\chi}\to V\ell^+)}
{\Gamma(\chi\to V\ell^-)+\Gamma(\bar{\chi}\to V\ell^+)}
\simeq \frac{2 \Re e({{\mathcal M}^{(0)}}^{\dagger}{\mathcal M}^{(1)})}
{{{\mathcal M}^{(0)}}^{\dagger}{\mathcal M}^{(0)}} \nonumber\\[4mm]
&&\hskip-30mm=\frac{4 m_{\chi}\{\Im m B_0(2 m^2_V+3 m^2_{\chi})+
\Im mC_0(2 m^4_V+2 m^2_V m^2_{\chi}+m^4_{\chi}+
m^2_h(m^2_V-m^2_{\chi}))\}}
{2 m^4_{\chi}-3 m^4_V+m^2_{\chi}m^2_V}\nonumber\\[4mm]
&&\hskip-30mm\times\ \frac{1}{(4\pi)^2}\frac{2 m^2_V}{v}
\ \frac{\Im m ({g^L_3}^{\ast}\ g_0 \ g^L_1)}{(|g^L_3|^2+|g^R_3|^2)}~,
\eaq
where
\baq{imbc}
\Im m B_0&=&\pi \frac{m^2_{\chi}-m^2_h}{m^2_{\chi}}
\ \Theta(m^2_{\chi}-m^2_h)~,\nonumber\\[4mm]
\Im m C_0&=&-\pi \frac{1}{m^2_{\chi}-m^2_h}
\log\biggl(1+\frac{(m^2_{\chi}-m^2_h)^2}{m^2_{\chi}m^2_V}\biggr)
\Theta(m^2_{\chi}-m^2_h)~,
\eaq
where $v=2 m_W/g$ and $\Theta$ denotes the step--function.
From the last line in Eq.~\rf{CPasym} we can see that 
the CP asymmetry in the case of the neutralino
decay is suppressed, since we have $|g^L_3| << |g^R_3|$.
For the chargino decay this relation is reversed, as
was mentioned above.

\end{document}